\documentclass[reprint,twocolumn,nofootinbib]{revtex4}
\usepackage[utf8]{inputenc}
\usepackage{graphicx}
\usepackage{url}
\usepackage{hyperref}
\usepackage{color}

\newcommand{\sNN}{s_\mathrm{NN}}
\begin{document}
\title{Jet overlap in heavy ion collisions at LHC energies and its consequences on jet shape}
\author{Iurii Karpenko$^{1,2}$, Joerg Aichelin$^1$, Pol Gossiaux$^1$, Martin Rohrmoser$^{3,4}$, Klaus Werner$^1$}
\affiliation{$^1$SUBATECH, Universit\'e de Nantes, IMT Atlantique, IN2P3/CNRS, \\
4 rue Alfred Kastler, 44307 Nantes cedex 3, France\\
$^2$Czech Technical University in Prague, FNSPE, B\v{r}ehov\'a 7, Prague 11519, Czech Republic\\
$^3$Institute of Physics, Jan Kochanowski University, 25-406 Kielce, Poland\\
$^4$H.~Niewodnicza\'nski Institute of Nuclear Physics PAN, 31-342 Cracow, Poland}

\date{\today}

\begin{abstract} \noindent
Central lead-lead collisions at the LHC energies may pose a particular challenge for jet identification as multiple jets are produced per each collision event. We simulate the jet evolution in central Pb-Pb events at $\sqrt{\sNN} = 2.76$~GeV collision energy with EPOS3 initial state, which typically contains multiple hard scatterings in each event. Therefore the partons from different jets have a significant chance to overlap in momentum space. We find that 30\% of the jets with $p_\perp > 50$~GeV, identified by the standard anti-$k_\perp$ jet finding algorithm with jet cone size R=0.3, contain `intruder' particles from overlapping generator-level jets. This fraction increases with increasing beam energy and increasing R. The reconstructed momentum of the jet differs from that of the modelled jet by the loss due to jet partons which are outside of the jet cone and by the gain due to intruder partons. The sum of both may be positive or negative. These intruder partons particularly affect the radial jet momentum distribution because they contribute mostly at large angles $\Delta r$ with respect to the jet centre. The study stresses the importance of the jet overlap effect emerging in central lead-lead collisions at the LHC energies, while being negligible in peripheral PbPb or $p$Pb/$pp$ collisions.
\end{abstract}


\maketitle

\section{Introduction}
Jets are created in hard collisions between elementary particles like quarks, gluons or $e^+e^-$. The leading jet particles escape from the reaction zone with a large transverse momentum and fragment into an almost collimated flow of hadrons, called jet cone. These jets can be well identified and these processes are well studied, theoretically as well as experimentally. If the hard collisions are embedded in an environment, as is it the case in high energy heavy-ion collisions, the situation is more complicated. The leading jet particle may interact with the environment, the plasma of quarks and gluons (QGP) \cite{Shuryak:1977ut}, what may lead to a transfer of energy and/or momentum to the QGP \cite{Gyulassy:1990ye}. Also the fragmentation of the leading jet particle into hadrons may be modified in the medium, especially for low energy hadrons which have a velocity equal or lower then the expansion velocity of the QGP. Both of these processes may change the distribution of hadrons in the jet cone and are presently under intensive study, theoretically \cite{Casalderrey-Solana:2015vaa, KunnawalkamElayavalli:2017hxo} as well as experimentally \cite{Sirunyan:2018qec, Aaboud:2019oac}.

Most of the hadrons created in heavy-ion reactions come the from QGP and their multiplicity is quite well described by statistical model calculations \cite{Becattini:2014hla, Sharma:2018jqf}. Therefore, they do not carry any information on the interaction among partons. Most of the jet partons do  not come to equilibrium with the partons from the QGP. The hadrons from jets fragmentation may therefore be
one of the few sources to obtain information about these parton-parton interaction. Jets are therefore very interesting objects and that is the reason for the present study.

One of the challenges of the study of jets in heavy-ion collisions is the identification of those hadrons which are produced by the fragmentation of the jet and to separate them from the background hadrons which are produced when the QGP hadronizes. For this task jet finding algorithms, which reconstructs jets by clustering the final state hadrons \cite{Dokshitzer:1997in, Wobisch:1998wt, Cacciari:2008gp}, have been advanced with the goal to provide the best possible approximation to a clean theoretical calculation of a solitary jet possibly modified by the medium, but without the medium itself. This is a very tedious task. In essence, the present paradigm is that the jet and hence all its properties (observables) are defined by a jet clustering algorithm, therefore the only way to perform an apple-to-apple comparison with experiment is to run the same jet finding algorithm over theoretically simulated events. This is feasible thanks to an open-source `industry standard' tool for the jet reconstruction, the FASTJET package \cite{Cacciari:2011ma} which incorporates most of the established algorithms for the jet reconstruction.

A basic observable in the jet physics which quantifies the medium modifications of jet properties in heavy ion collisions is a jet nuclear modification factor $R_{AA}(p_T)$ \cite{Khachatryan:2016jfl, Adam:2015ewa} which is the ratio of the jets in a given transverse momentum $p_\perp$ bin observed in AA collisions to that observed in pp collisions properly weighted by the number of elementary pp collisions one expects (for a given centrality) in heavy-ion collisions. If a heavy ion collision is only a superposition of pp collisions and if jets are properly identified we expect $R_{AA}=1$ and any deviation from this values signals either that the jet creating partons have a different distribution in AA as compared to pp or that the jet is modified during its passage through the QGP. However, in heavy ion collisions - especially in central heavy ion events - not only a medium is created which modifies the jet properties and jet evolution, but also multiple hard scatterings happen in each event. Therefore the final states of the jets have significant chances to overlap in momentum space.

Another observable of interest is a jet shape, which actually consists of many different observables quantifying different aspects of the substructure of a jet \cite{Acharya:2018uvf}. Here we will be particularly interested in a radial momentum distribution (sometimes referred to as `jet shape' by CMS \cite{Sirunyan:2018jqr}) which shows how the overall jet momentum is distributed in the jet cone, characterised by the jet radius $R$. Without medium effects and assuming a correct reconstruction of the jets it should be identical in $pp$ and PbPb. ALICE and CMS collaborations recently studied the jet shape in PbPb and its modification with respect to the $pp$ baseline at $\sqrt{\sNN}=2.76$ and 5.02 TeV, respectively. CMS reports on the shift of the momentum away from the jet axis out to large relative angular distance \cite{Sirunyan:2018jqr}, which is supported by ALICE results \cite{Acharya:2019ssy}, however the error bars in the latter case are too large to provide a solid observation of the jet shape modification in PbPb.

The question how a partial jet overlap can affect the two above-mentioned observables - the jet momentum and the jet shape - is the topic of this article. In the present paper we investigate this aspect of the jet reconstruction with the widely used jet finding code FASTJET using the anti-$k_t$ algorithm for jets in PbPb collisions at the LHC energies simulated with EPOS3-Jet framework.
We describe our modelling framework in Section~\ref{sec:model}, introduce the concept of `jet purity' in our modelling, which quantifies the magnitude of the jet overlap, and show the effects of jet overlap on the selected jet observables in Section~\ref{sec:results} and conclude in Section~\ref{sec:conclusions}.

\section{Model}\label{sec:model}
\textbf{Initial state.} An essential point of the study is that we use EPOS3 model \cite{Werner:2013tya} (version 3.238) to generate both hydrodynamic initial state and initial hard partons - seeds from which the jets develop. Each elementary $NN$ interaction in the EPOS approach is treated in the Gribov-Regge multiple scattering framework. Each individual scattering is referred to as a Pomeron, represented by a parton ladder \cite{Drescher:2000ha,Werner:2007bf,Werner:2010aa}. The initial hard partons are produced in the hardest process in each ladder, which can be identified as Born process.

\begin{figure}
    \centering
    \includegraphics[width=0.5\textwidth]{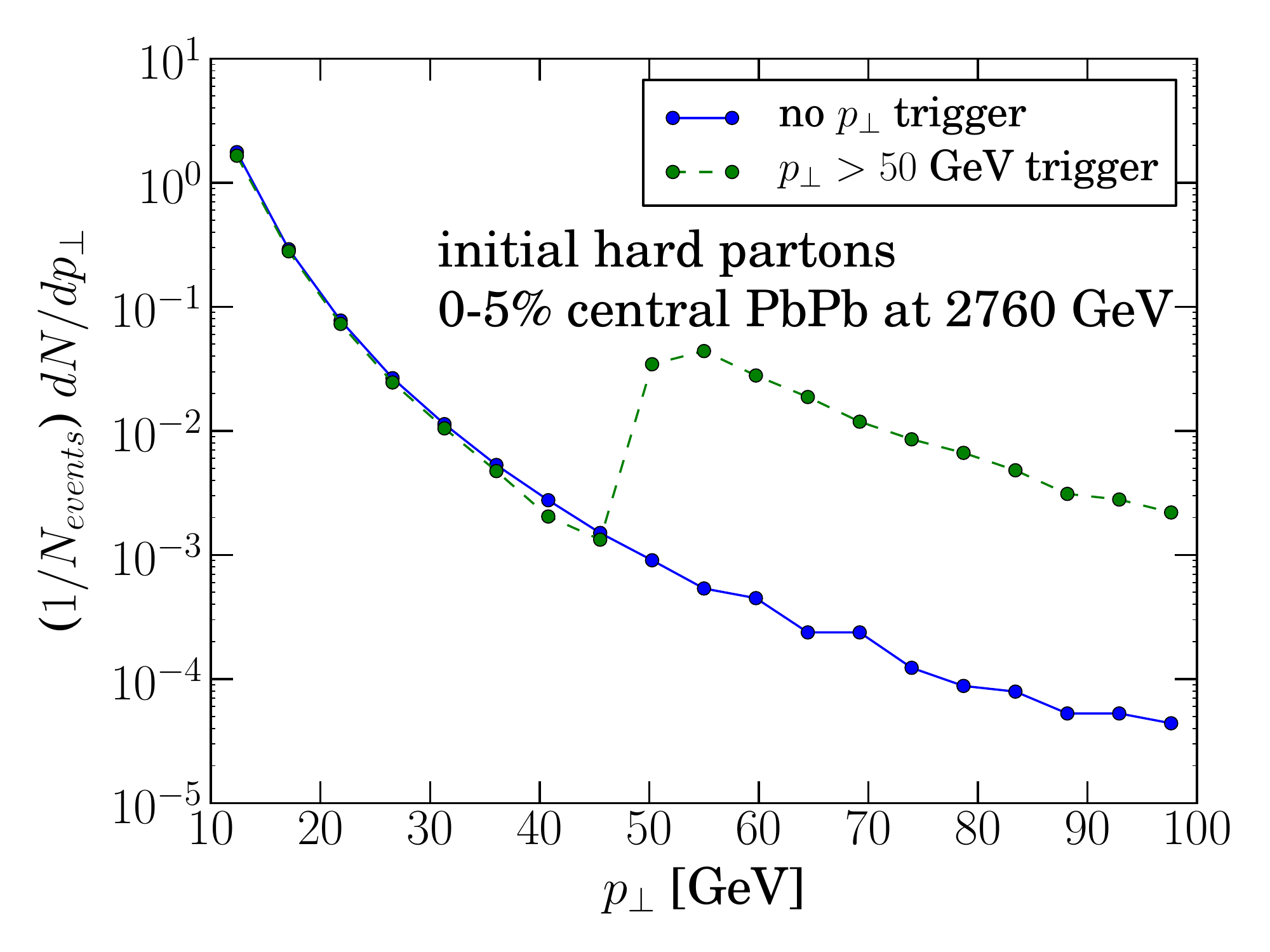}
    \caption{Transverse momentum distribution of hard partons produced within the rapidity interval $|y|<2$ in the initial state EPOS calculations of 0-5\% central PbPb collisions at $\sqrt{\sNN}=2.76$~TeV. Solid curve corresponds to untriggered events, whereas dashed curve corresponds to the trigger of at least one parton with $p_\perp>50$~GeV in an initial state configuration.}
    \label{fig:iniHardPartons}
\end{figure}

The resulting transverse momentum spectrum of initial hard partons extends down to relatively low $p_\perp$ of a few GeV. As an example we show in Fig.~\ref{fig:iniHardPartons}  such transverse momentum spectrum for Pb-Pb collisions at $\sqrt{\sNN}=2.76$~TeV energy (solid curve) . As one can see, in each central Pb-Pb collision at the LHC energy there are, within 4 units of rapidity, of the order of ten hard partons with $p_\perp>10$~GeV .

Experimental analyses usually set a $p_\perp$ trigger for the jet studies, e.g.\ one selects only the events with at least one jet with $p_\perp$ larger than a certain value, typically 50-80 GeV. We simulate such $p_\perp>50$~GeV trigger in our initial state calculations by setting a corresponding trigger to have at least one initial hard parton with $p_\perp>50$~GeV in an event, which results in the dashed curve on Fig.~\ref{fig:iniHardPartons}. Unsurprisingly, the triggered events are still dominated by the partons with lower $p_\perp$.

\textbf{Jet evolution.} The initial hard partons serve as seeds for jet evolution, which is performed with a Monte Carlo implementation \cite{Rohrmoser:2018fkf} of DGLAP equations. The initial virtuality scale of the hard partons is set as $Q^2_{\rm ini}=p_\perp^2$ and the evolution of the jet (parton splittings) proceeds until all of the partons reach a lower virtuality scale $Q_0=0.3$~GeV. For the present study we do not apply a hadronization procedure to the final state of the jet, therefore what is discussed and presented below is on the \textit{parton level}.

Also, for the purpose of present study all the medium effects are switched off, therefore the jets modelled in PbPb collisions do not have extra broadening - we call them $pp$-like jets. We made the latter simplification for the purpose of clarity of the message of the paper, since a particular form of the in-medium energy loss is not essential for the effects under discussion.

\textbf{Jet reconstruction.} The ensemble of partons coming from the fragmentation of one single jet particle we called {\it modelled jet}. The modelled jet is a purely theoretical quantity but well defined because we can follow the trajectory of all particles in the simulations. The modelled jet has to be confronted with the {\it reconstructed jet} or, in short, {\it jet} which is the ensemble of partons which a jet finding algorithm has identified as coming from the same jet. This is a quantity which can be compared to experiment.
The final parton momenta of all modelled jets in a single event (partons at the virtuality scale $Q_0$), are transferred to FASTJET 3.3 \cite{Cacciari:2011ma} in order to perform the jet finding using the anti-$k_T$ algorithm. The QGP hadrons are not generated, therefore FASTJET is dealing with jet partons only. We make this approximation to avoid additional complication due to background subtraction or due to different hadronization schemes which have nothing to do with the message of this paper. FASTJET then returns the list of reconstructed jets for each event. 

\section{Results and discussion}\label{sec:results}

\begin{figure}
    \centering
    \includegraphics[width=0.5\textwidth]{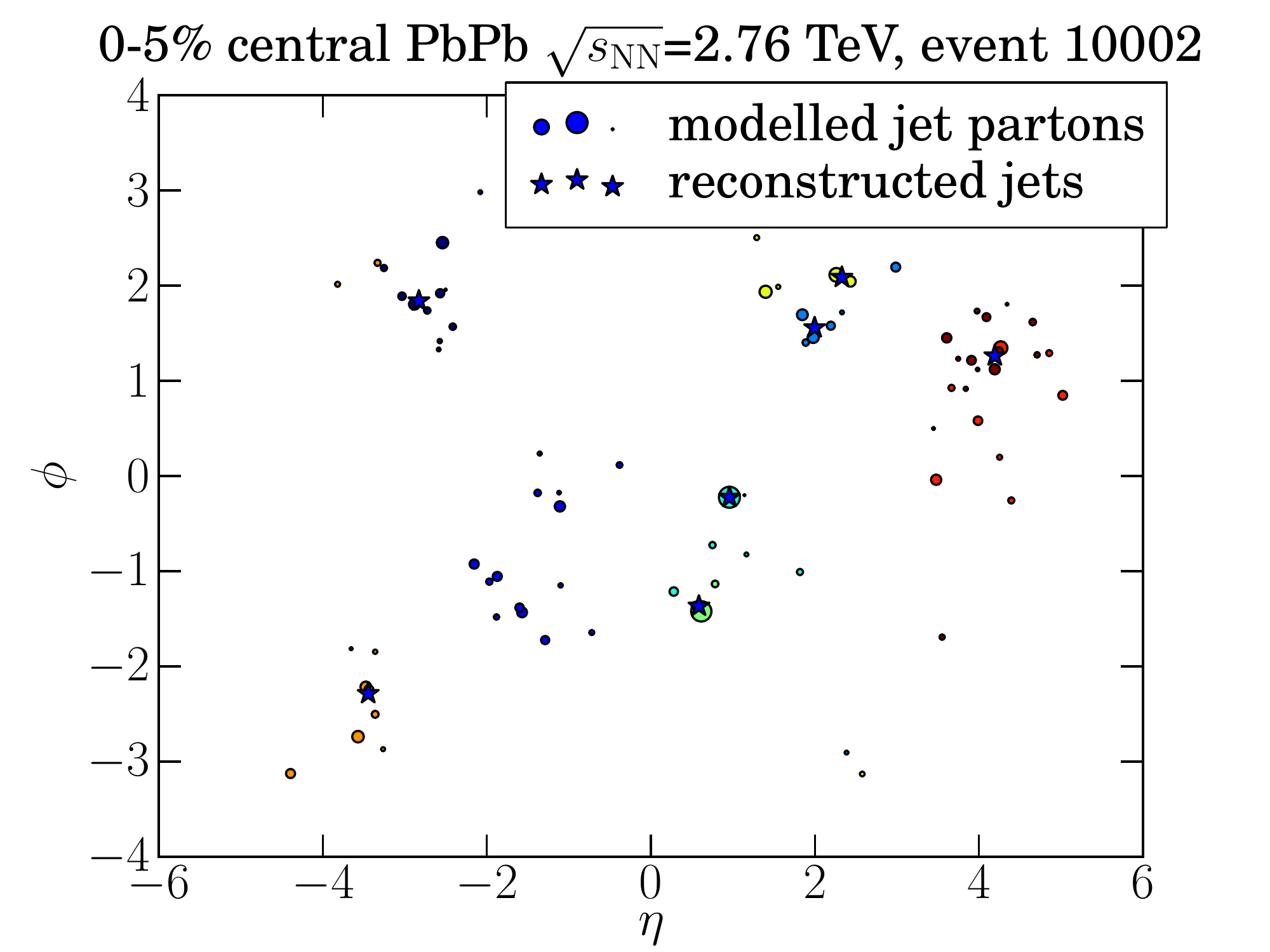}
    \caption{Distribution of final state jet partons (circles) and reconstructed jets (stars) in the $\eta-\phi$ plane in a randomly picked central PbPb event at $\sqrt{\sNN}=2760$~GeV collision energy, modelled with EPOS initial state.}
    \label{fig:jets-visualised}
\end{figure}

In Fig.~\ref{fig:jets-visualised} we display a randomly chosen jet event corresponding to central PbPb collision at the $\sqrt{\sNN}=$2.76 TeV energy. In this figure both, final state partons from the modelled jets (circles) and of the reconstructed jets (stars) are plotted in the $\eta-\phi$ plane, where $\eta$ is pseudorapidity and $\phi$ is the azimuthal angle of a jet parton or a reconstructed jet. For the modelled jets, the symbol size of the partons is proportional to its momentum and partons from the same modelled jet have the same colour. One can see that in the most cases FASTJET  identifies the jets correctly by calculating the location of each reconstructed jet around the `center of momentum' of the final state of the modelled jet. However, two artefacts may happen: (i) some of the final state partons are too far from the core of the jet, to be clustered together with the rest of the partons into a corresponding (reconstructed) jet. This reduces the momentum of the jet as compared to the modelled jet. (ii) The jet partons from different modelled jets may happen to be so close that they overlap in the $\eta-\phi$ space, and the jet finding algorithm clusters them together. In the case the momentum of the jet increases. Consequently, both effect can balance each other. How often do those happen actually?

It is rather straightforward to quantify these effects of the jet finding algorithm with our simulated events. We
know to which modelled jet and to which reconstructed jet each parton belongs by passing the jet index of each modelled jet parton to FASTJET via an auxiliary variable \texttt{user\_index}, and using this variable in the FASTJET output.

The number of partons in a jet is not an infrared safe observable. Therefore to quantify the effect we evaluate the fractions of momenta from different modelled jets in each reconstructed jet. One expects that each reconstructed jet contains a dominant momentum fraction from a particular underlying modelled jet plus smaller fraction(s) from the neighbouring modelled jets which partially overlap in momentum space.

For each reconstructed jet we calculate which fraction of its total momentum comes from which modelled jet. This way we can also see which modelled jet is dominant for each reconstructed jet.

\begin{figure}
 \includegraphics[width=0.5\textwidth]{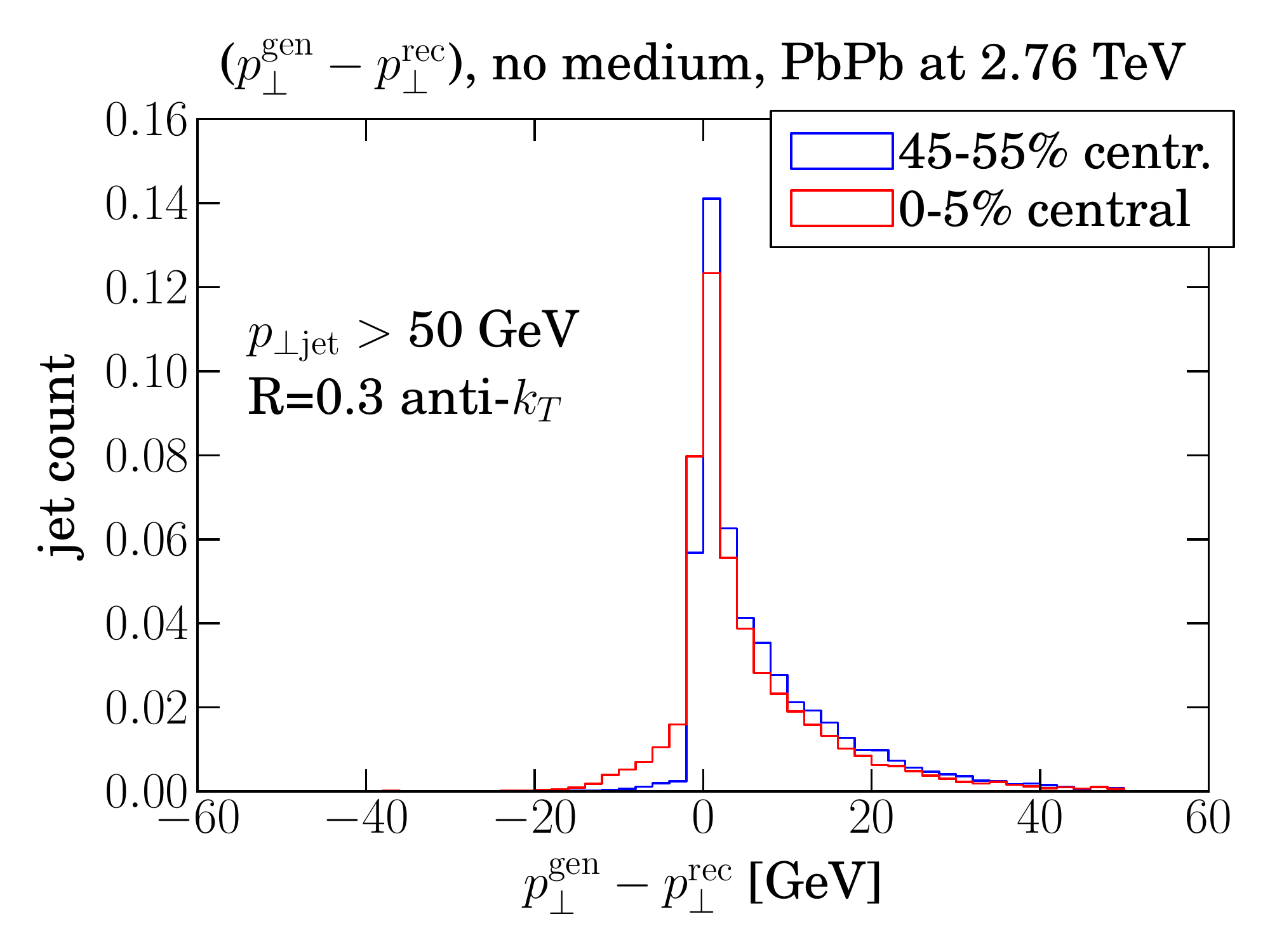}
 \caption{Distribution of the difference between the modelled jet $p_\perp$ and reconstructed jet $p_\perp$ in 45-55\% and 0-5\% central Pb-Pb collisions at $\sqrt{\sNN}=2.76$~TeV LHC energy.}
 \label{fig:deltaPt}
\end{figure}

The two artefacts of jet reconstruction discussed in the beginning of the section are clearly observed on Fig.~\ref{fig:deltaPt} where we plot the distribution of difference between the dominant modelled jet and reconstructed jet. There one can see that indeed in most of the cases the difference is positive - meaning that the reconstructed jet momentum is smaller than the momentum of the corresponding modelled jet, and this part of the distribution does not depend much on the centrality of collision. This happens when the anti-$k_\perp$ algorithm leaves some of the modelled partons out, because they are too far from the rest in the $\eta-\phi$ space. The latter is a very well known feature of the jet finding procedure. There is however also a part of the distribution where the difference is negative - meaning that the reconstructed $p_\perp$ is larger. Interestingly one can see that the negative part is much more pronounced for 0-5\% central case than for 45-55\% central. This part corresponds to cases when the jet finding algorithm clusters together partons from neighbouring jets, even compensating for the left out partons from the corresponding modelled jet.

\begin{figure}
 \includegraphics[width=0.5\textwidth]{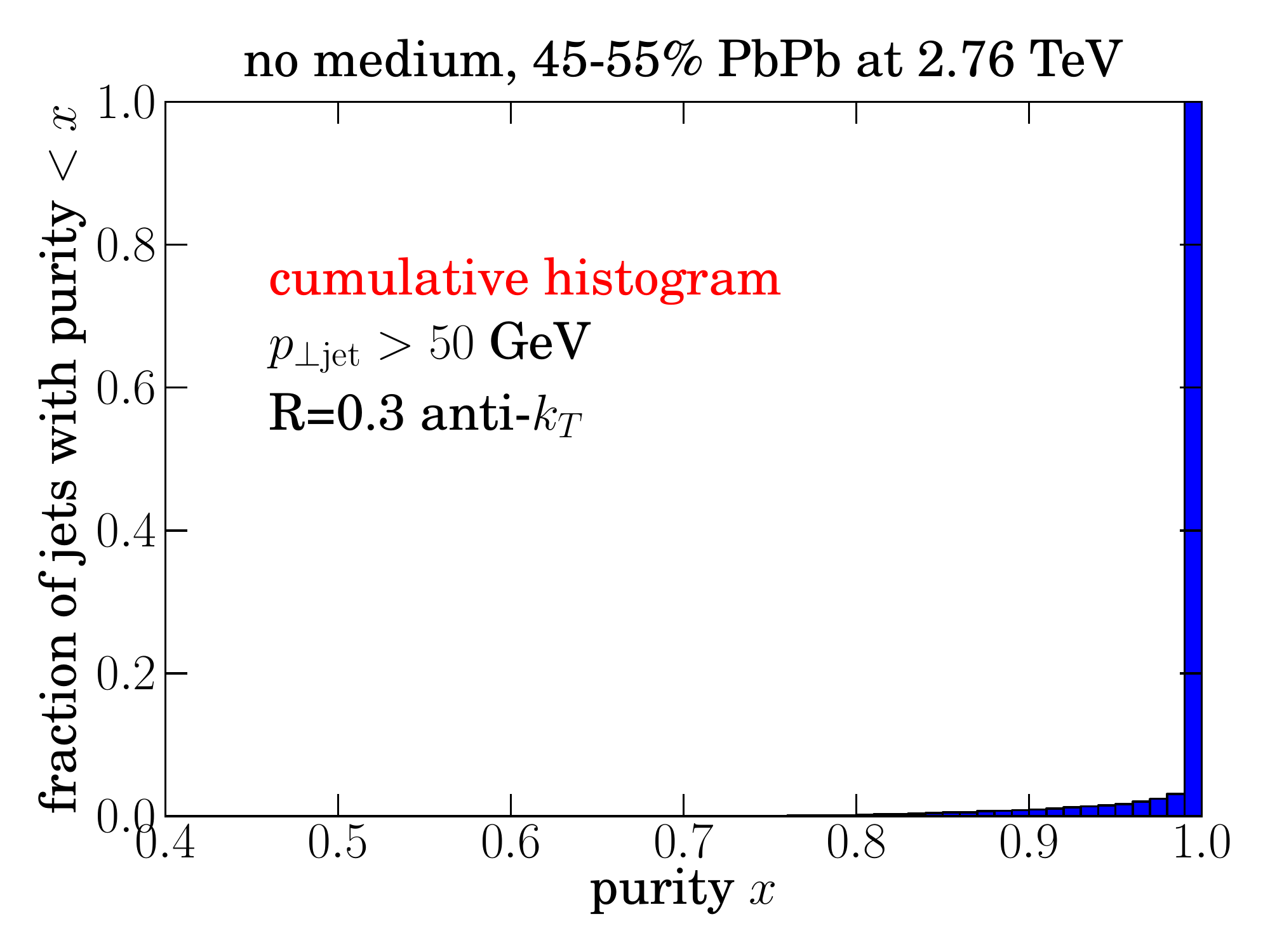}
 \caption{Cumulative distribution of purity of the jets in 45-55\% central PbPb collisions at $\sqrt{\sNN}=2.76$~TeV LHC energy modelled with EPOS3 initial state.}
 \label{fig:purityNoncentral}
\end{figure}

In what follows we will focus on the second artefact of the jet reconstruction - clustering together partons from neighbouring jets. We define a quantity which we call a `\textit{jet purity}', which is a fraction of the momentum of the reconstructed jet coming from the corresponding (dominant) modelled jet. The other fractions of the reconstructed jet momentum come from neighbouring modelled jets. On Fig.~\ref{fig:purityNoncentral} we show a cumulative distribution of the jet purity for the simulations of 45-55\% central PbPb events at the $\sqrt{\sNN}=2.76$~TeV energy. The distribution shows the faction of jets which have purity equal or lower than $x$ when reconstructed. The distribution is strongly peaked around 1, which means that almost all of the jets, reconstructed by FASTJET, have practically all of their momentum coming from a single underlying modelled jet. Here the jet finding works quite well. From the cumulative distribution one can conclude that only around 3\% of the jets have their purity somewhat different from 1, which means that for around 97\% of cases there is no ``momentum contamination'' from neighbouring jets in the reconstruction procedure.

\begin{figure}
 \includegraphics[width=0.5\textwidth]{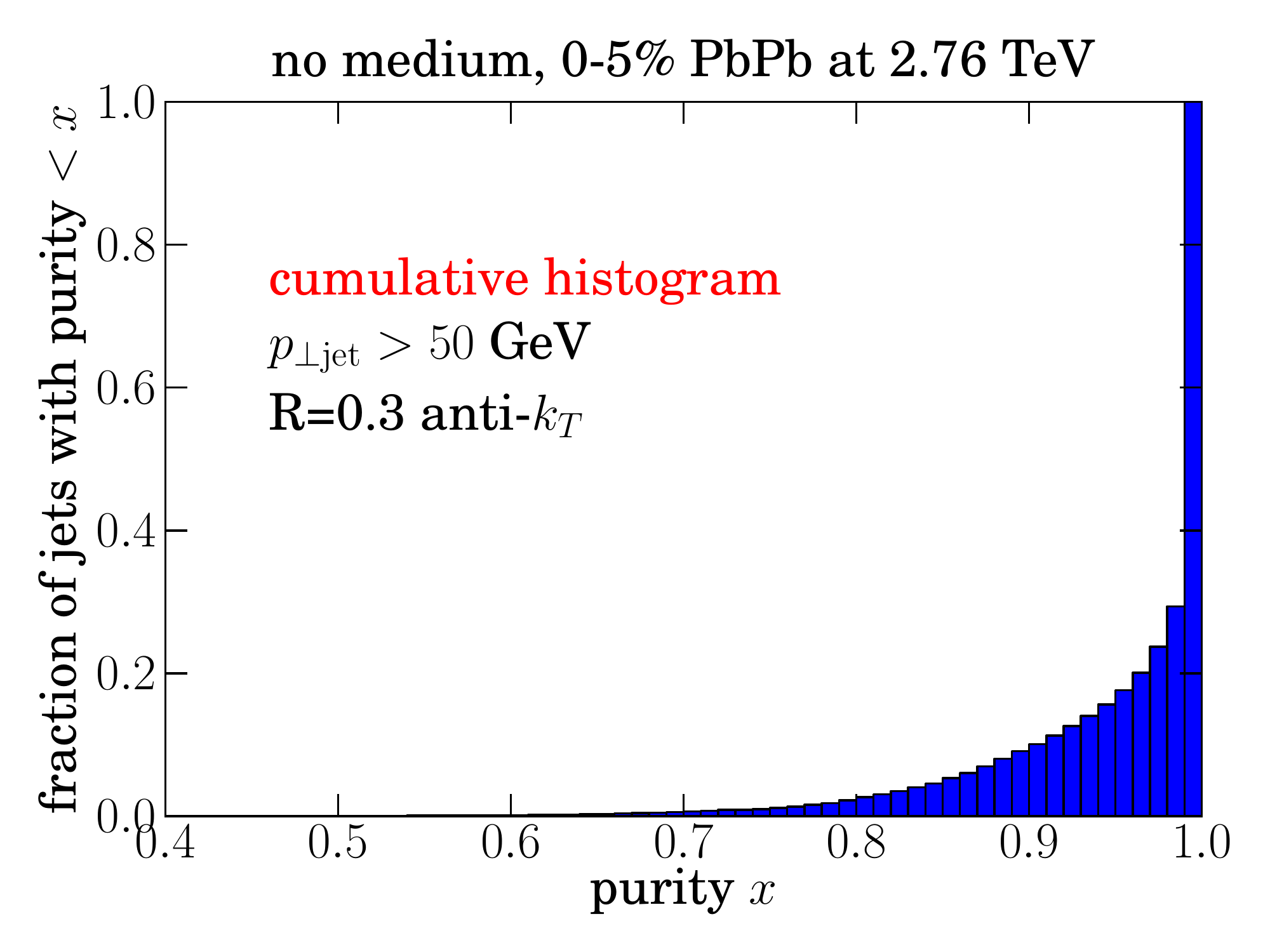}
 \caption{Same as Fig.~\ref{fig:purityNoncentral} but for 0-5\% central PbPb collisions.}
 \label{fig:purityCentral}
\end{figure}
 
We move on to jets created in the most central (0-5\% centrality) events at the same energy. Fig.~\ref{fig:purityCentral} shows that the contamination increases with centrality. In this case around 30\% of the jets have a ``momentum contamination'' from the neighbouring jets ranging from a small fraction to around 30\% of their total momenta.

\begin{figure}
 \includegraphics[width=0.5\textwidth]{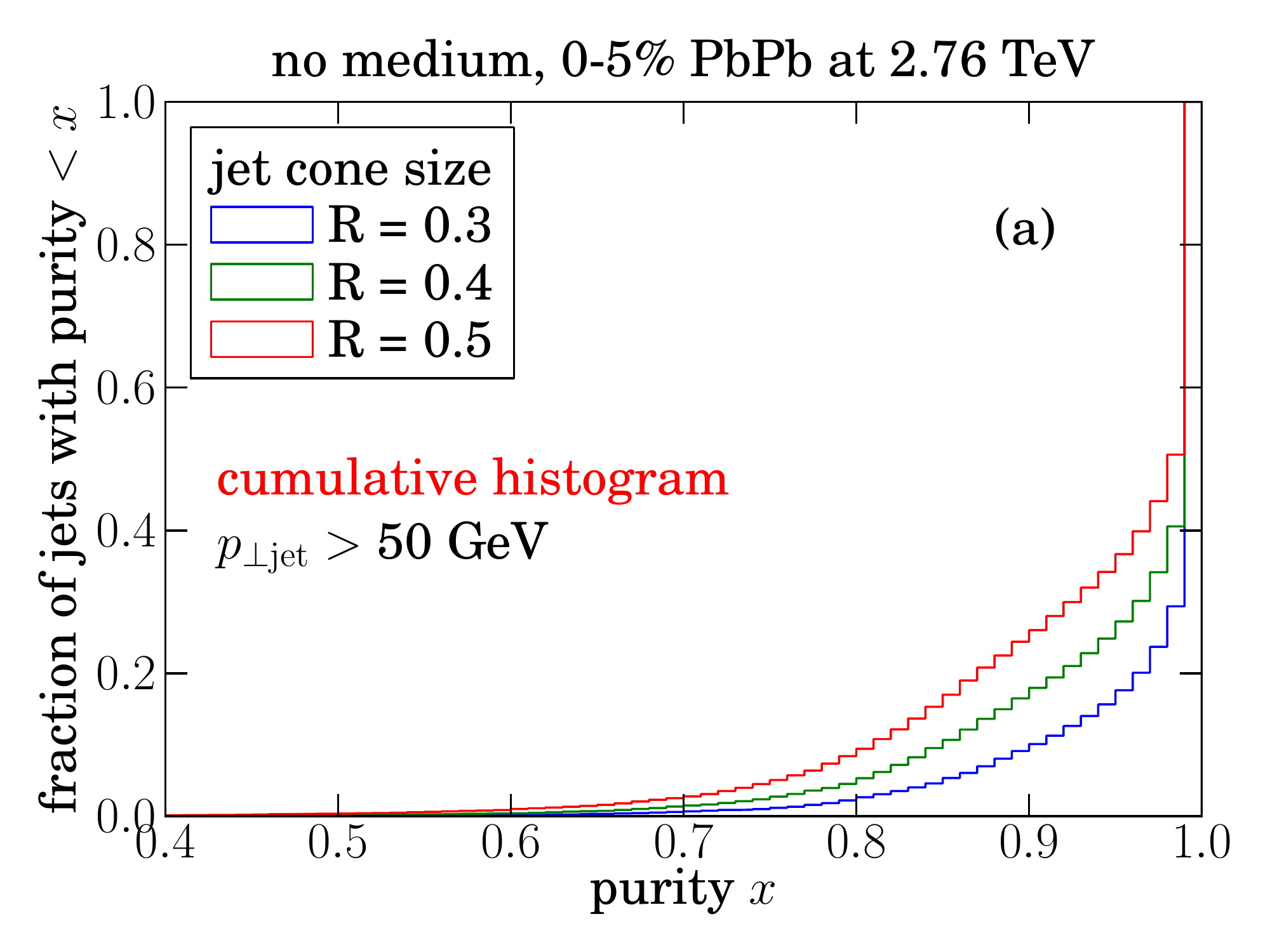}
 \includegraphics[width=0.5\textwidth]{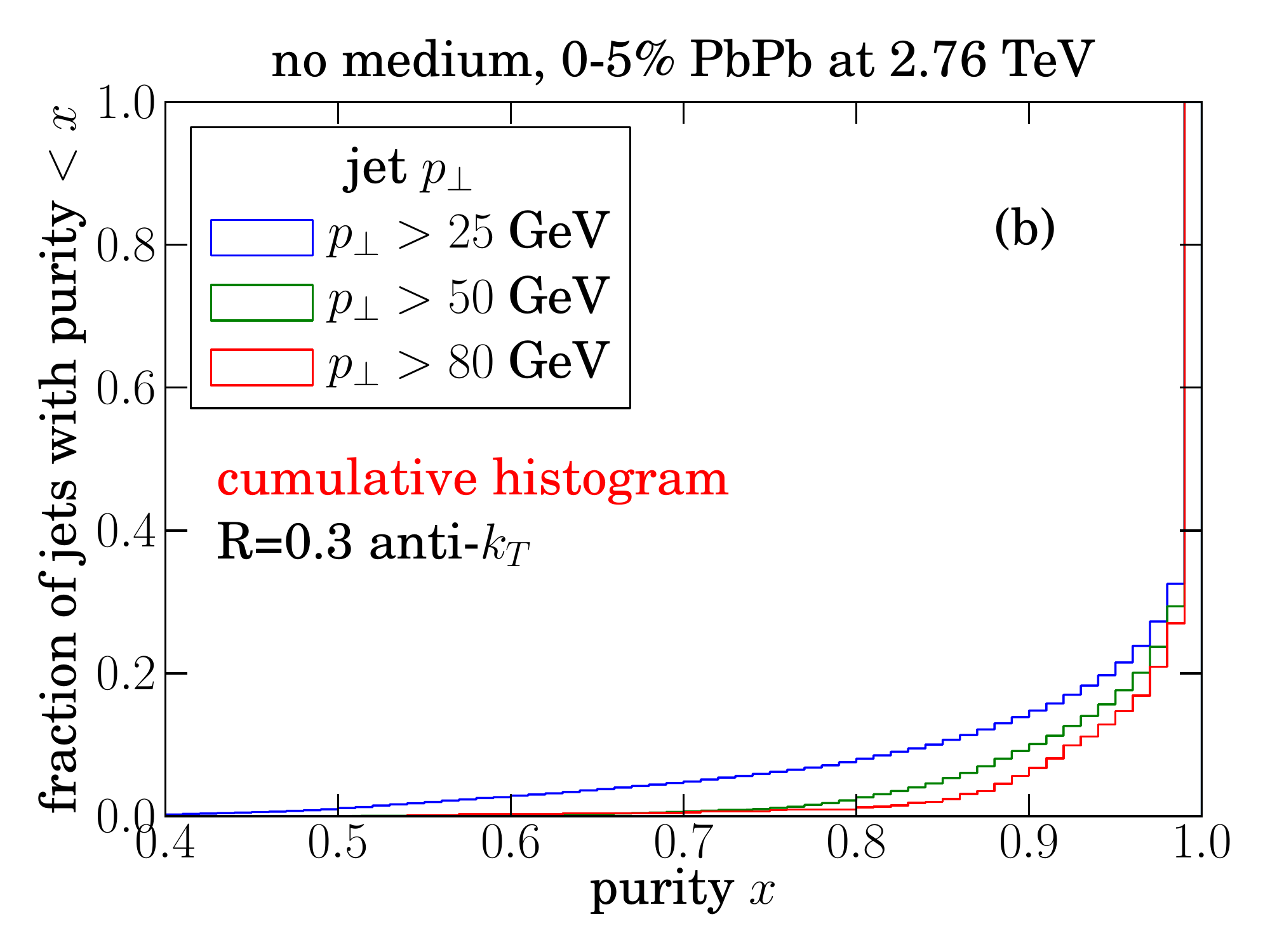}
 \caption{(a) jet $p_\perp$ dependence and (b) jet cone size dependence of the jet purity.}
 \label{fig:purityEffects1}
\end{figure}

Both, the jet $p_\perp$ selection as well as the jet cone size $R$ in the jet finding algorithm, influence strongly  the magnitude of the contamination effect, as one can see from Fig.~\ref{fig:purityEffects1}. On the top panel this is displayed for fixed jet $p_\perp$ threshold as a function of the cone size, whereas the bottom panel corresponds to fixed jet cone size $R$ in the jet reconstruction algorithm and different $p_\perp$ selection of the jets. Increasing the jet cone size from $R=0.3$ to $R=0.5$ reduces to 50\%  the fraction of pure (uncontaminated) jets. This is expected as with larger jet cone size the chances to cluster together the neighbouring jets increase.

We come now to the second jet observable which is employed to compare jets created in pp with those created in PbPb, the radial distribution of momentum within the jet cone with respect to the jet axis.
For the jet shape analysis one sums over the $\Delta \eta, \Delta \phi$ bins which have a distance $r= \sqrt{(\Delta \eta)^2+ (\Delta \phi)^2}$ from the jet axis. There are slightly different definitions of the jet shape found in the literature, thus here we follow the CMS definition:
\begin{equation}
 \rho(r) = \frac{1}{p_\perp^{\rm jet}} \frac{1}{2\delta r} \sum_{{\rm track} \in (r-\delta r, r+\delta r)} p_\perp^{\rm trk}
\end{equation}
where the momenta of the tracks (hadron, or in our case parton) are summed up for each bin [$(r-\delta r, r+\delta r)$].

\begin{figure}
 \centering
 \includegraphics[width=0.53\textwidth]{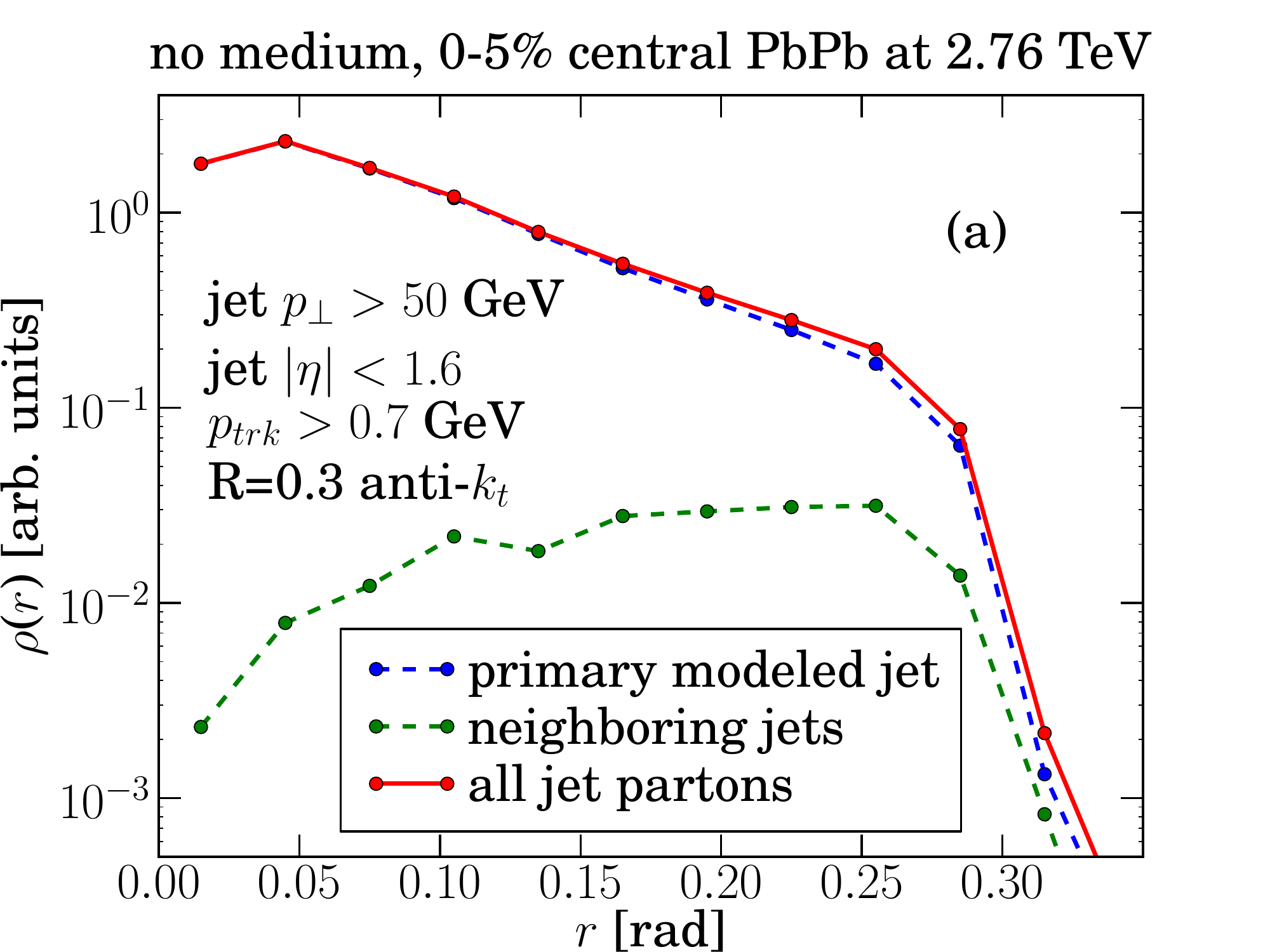}\\
 \includegraphics[width=0.53\textwidth]{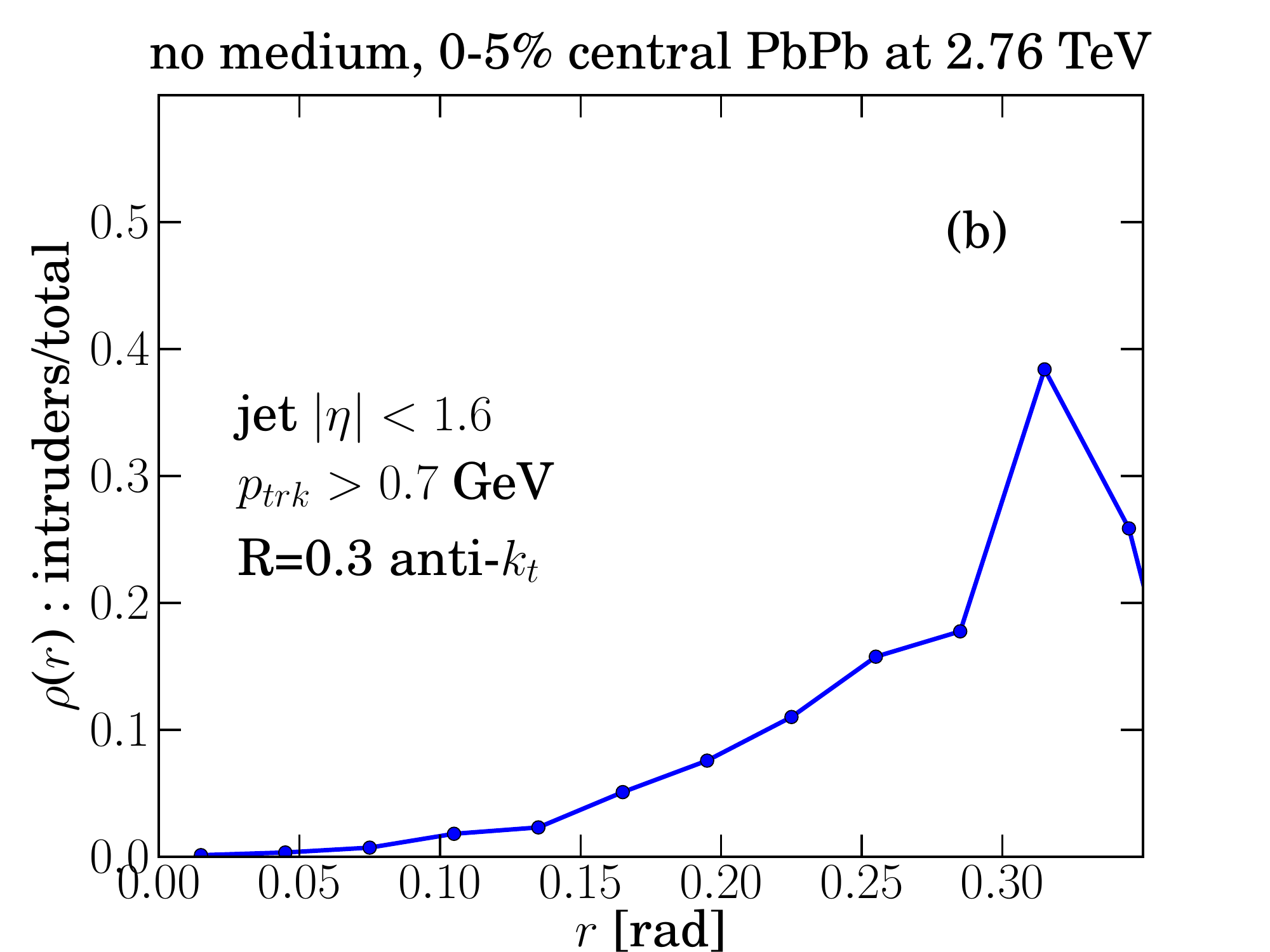}
 \caption{(a) jet shape $\rho(r)$ of reconstructed jets in 0-5\% central PbPb collisions. Solid red curve corresponds to the full jet, whereas dashed curves correspond to contributions from corresponding modelled jets (blue) and neighbouring jets (green). (b) fraction of the neighbouring jet contribution in the total jet shape.}
 \label{fig:jetRho}
\end{figure}

\begin{figure}
 \centering
 \includegraphics[width=0.53\textwidth]{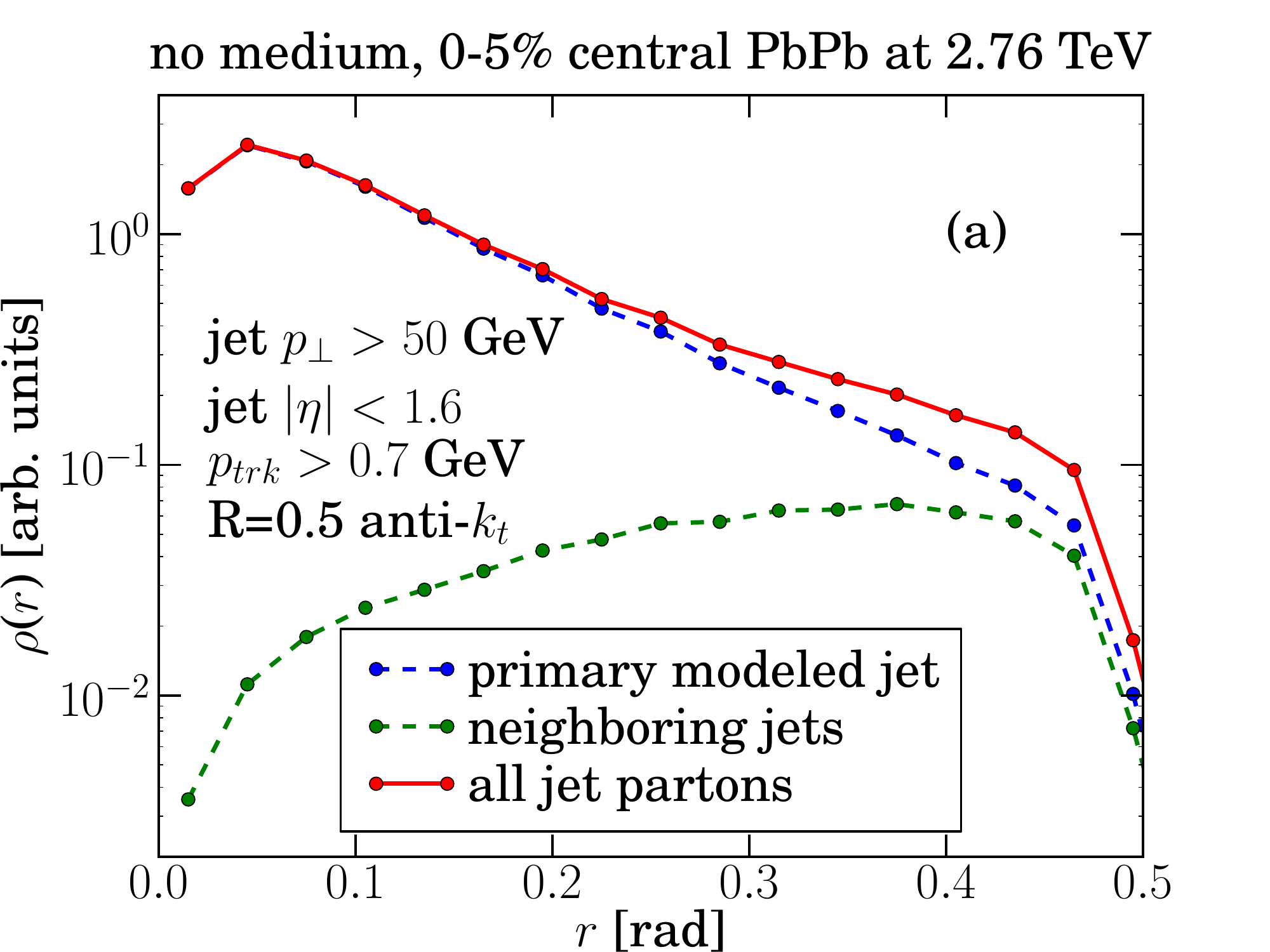}\\
 \includegraphics[width=0.53\textwidth]{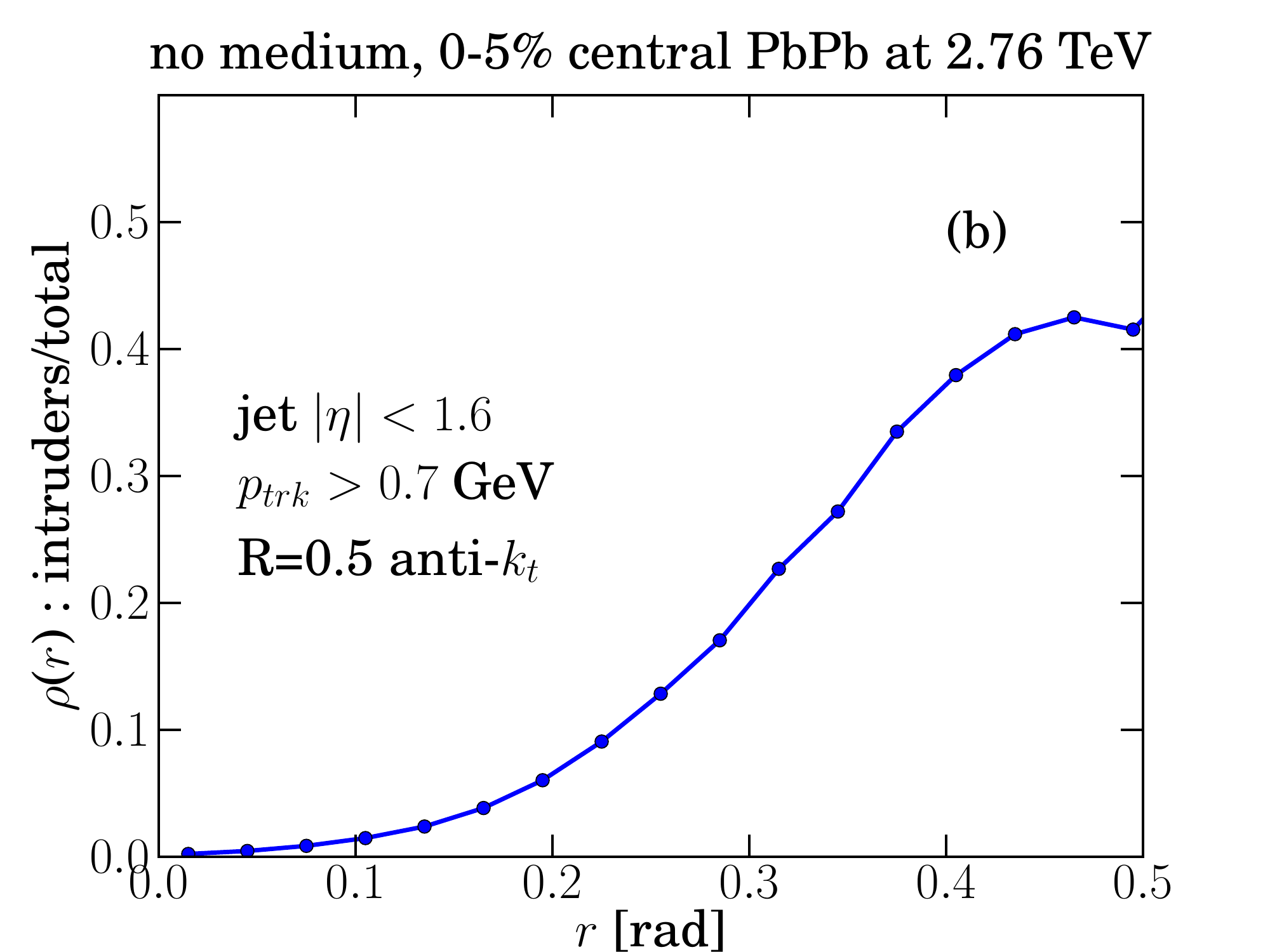}
 \caption{Same as Fig.~\ref{fig:jetRho} but for jets reconstructed with $R=0.5$ setting.}
 \label{fig:jetRho_R05}
\end{figure}

In Figs.~\ref{fig:jetRho} and ~\ref{fig:jetRho_R05}, top, we show contributions of the dominant modelled jet and neighbouring jets (intruders) to the momentum in a given $r$ bin as well as the sum of the two, as a function of angle $r$ with respect to the jet axis. In the bottom panels we show the relative contribution of the intruder jets as a function of this angle. For these plots we have selected jets with $p_\perp>50$~GeV.
In Fig.~\ref{fig:jetRho} one can see that indeed the intruder jets contribute negligibly to the core (small $r$) of the reconstructed jet, whereas at the periphery (larger $r$) the intruder jets start to be non-negligible. From the bottom panel of Fig.~\ref{fig:jetRho} one can see that the neighbouring jets contribute with almost 20\% at $r=0.25$ to the momentum of the reconstructed jet. If the jet cone size is increased to $R=0.5$, Fig.~\ref{fig:jetRho_R05}, the intruder jet contribution stays the same at given $r$ as for $R=0.3$, but now the jet cone extends to larger $r$ values - and the relative contribution of the intruder jets raises further with $r$ to above 40\%.

\begin{figure}
 \centering
 \includegraphics[width=0.53\textwidth]{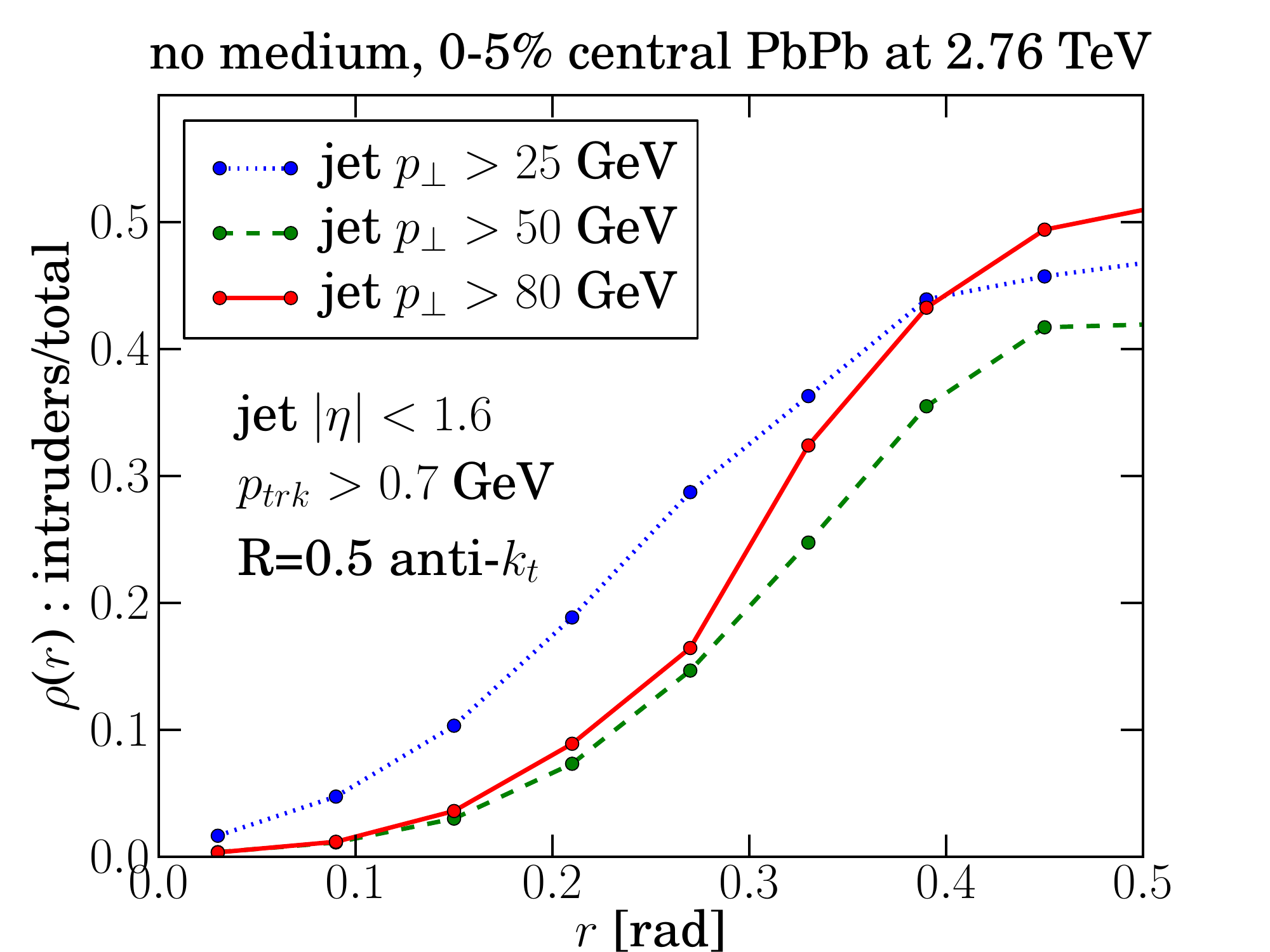}
 \caption{Jet $p_\perp$ dependence of the momentum fraction from the neighbouring jet, for the jets reconstructed with $R=0.5$ setting.}
 \label{fig:jetRho_ptDep}
\end{figure}

Finally, in Fig.~\ref{fig:jetRho_ptDep} we show the relative contribution from intruder jets to the jet shape as a function of $r$ for different total momentum $p_\perp$ of the jet. Similarly to the previous plots, the contribution of the intruder jets to the total momentum increases with the opening angle, and one may note that it depends only weakly on the jet $p_\perp$. The latter is again a consequence of the fact that most of the $p_\perp$ of the jet is localised at small cone angles (core of the jet), whereas the angular density of momentum in the jet corona (at large $r$) does not increase linearly with the $p_\perp$ of the jet - which keeps the contribution from the intruder jets significant, independent of $p_\perp$.

As the jet overlap is negligible in peripheral PbPb events, and consequently in $p$Pb and $pp$ collisions, the contribution of the intruder jets shown on Fig.~\ref{fig:jetRho_ptDep} corresponds quantitatively  to the relative modification of jet shape in central PbPb collisions as compared to the one in $pp$.

\begin{figure}
    \centering
    \includegraphics[width=0.53\textwidth]{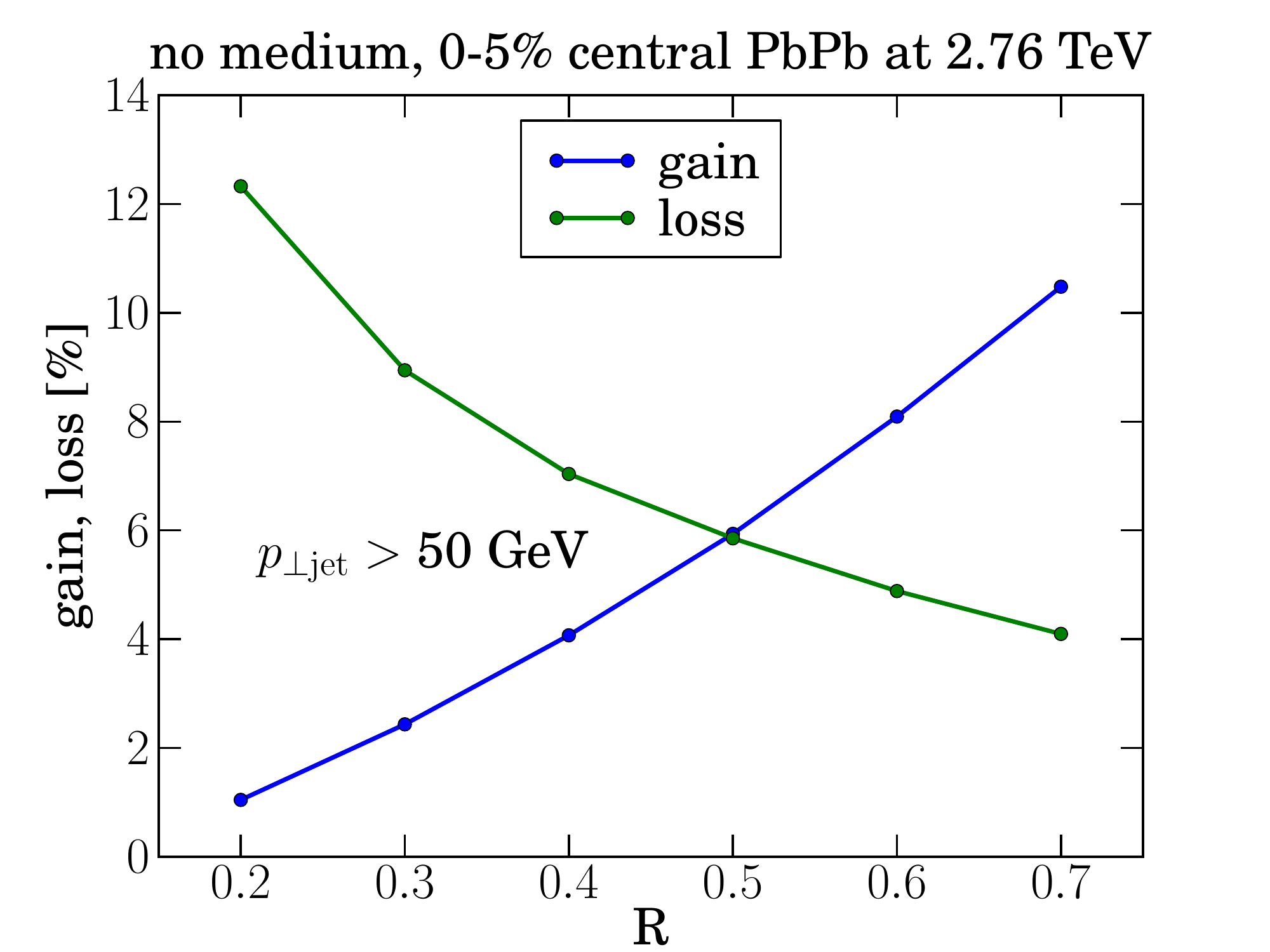}
    \caption{Fraction of momentum of modelled jet not clustered into a corresponding reconstructed jet (loss) and a fraction of momentum from the intruders (neighbouring jets) in the reconstructed jet (gain), both as a function of the jet cone size R in the jet reconstruction.}
    \label{fig:gain_loss}
\end{figure}
Therefore, as far as central heavy-ion collisions at the LHC energies are concerned, there seems to be no sweet spot for the jet cone size $R$ in the jet reconstruction procedure. The smaller we chose the value of R the less numerous are the intruder partons from other jets but the more partons from the jet of the interest are `lost' (clustered into other jets) in the reconstruction. We demonstrate it on Fig.~\ref{fig:gain_loss} where we plot gain and loss fractions which we define as:
$${\rm gain}=p_\perp^{\rm intruders}/p_\perp^{\rm reconstructed}$$
$${\rm loss}=p_\perp^{\rm not-clustered}/p_\perp^{\rm modelled},$$
where $p_\perp^{\rm modelled}$ is the momentum of the generator-level (modelled) jet, $p_\perp^{\rm not-clustered}$ is the fraction of the momentum of the modelled jet which is not clustered into the corresponding reconstructed jet. The gain fraction is essentially equal to $(1-x)$, where $x$ is purity, as the total jet momentum is composed from the corresponding modelled jet and the intruder partons. In case of peripheral Pb-Pb (and by extension in p-Pb or pp) collisions the effects of jet overlap become small, therefore the gain term will drop down considerably whereas the loss term will stay on the same level. Therefore choosing a larger value for $R$ for the jet reconstruction seems to be a safe option for peripheral PbPb collisions.

Generally speaking, the presented results demonstrate the importance of the jet overlap effect for jet reconstruction in a busy heavy ion environment, both from experimental and from theoretical side. In the experimental analysis, the overlap effect is considered as a part of background, or underlying event (UE), the latter one being dominated by low-$p_\perp$ hadrons. However one should keep in mind that the background subtraction is performed on the observable-by-observable basis and the techniques differ from experiment to experiment and from analysis to analysis. Intricate observables such as the radial momentum distribution $\rho(r)$ shown here, seems to be more tedious to thoroughly correct for in the experiment. For the radial jet momentum distribution CMS reports to perform a background subtraction in a statistical way based on PYTHIA+HYDJET simulations \cite{Sirunyan:2018jqr} - which also removes the jet overlap effects. Somewhat different from that, in \cite{Acharya:2019ssy} ALICE reports the ratio of actual jet shape in PbPb events relative to the shape of PYTHIA jets embedded into actual PbPb events, as a proxy for the PbPb/$pp$ ratio\footnote{However, the resulting PbPb/embedded ratio, which should only contain a clear medium recoil effect, has large error bars which extend down to the value 1.0. This does not allow to make a solid conclusion that the radial momentum distribution is modified in PbPb with respect to $pp$.}. To our opinion the latter is a more direct quantification of the jet shape modification in PbPb, provided that PYTHIA describes the jet shape in $pp$ accurately enough. For the measurement of the jet mass - which is another observable sensitive to the jet substructure - ALICE subtracts the background for each jet with the area-based subtraction method \cite{Acharya:2017goa}, which presumably subtracts only the soft background. For the measurement of yields of charged particles $D(p_\perp,r)$ within a jet, ATLAS \cite{Aad:2019igg} performs a background subtraction based on the transverse energy density in the $\eta-\phi$ regions not populated by jets and then rejects jets which are located within an angular distance $\Delta r<1.0$ of each other. ATLAS claims the latter requirement rejects as little as 0.01\% of jets, presumably because all jets are required to have $p_\perp>126$~GeV.

As one can see, the techniques for the background subtraction (including jet overlap subtraction) are not standardised, new techniques such as machine learning are coming up and the goal is to provide a cleansed experimental result which can be compared to a theoretical calculation of a solitary jet passing through the dense QGP medium.

From the theory side, there are two ways the jet overlap effect can be avoided when comparing calculations to the experimental data: either (1) a theory calculation models solitary jets passing through the QGP medium such as in \cite{Tachibana:2017syd}, \cite{KunnawalkamElayavalli:2017hxo} or (2) full AA event with multiple jets is modelled such as in \cite{Putschke:2019yrg}, \cite{He:2018xjv} - then one also has to include the soft background and perform a background subtraction procedure which matches an experimental one, such as described in \cite{He:2018xjv}. Case (1) by construction does not introduce the jet overlap effect, whereas in case (2) the jet overlap effect is removed similarly to experiment. We argue that in case (2) it is not enough to exclude the low-momentum hadrons of non-jet origin (e.g.~from hydrodynamic approach) and apply the jet finding procedure over the combined final states of the jets, because the jet overlap effect will emerge in the reconstructed theory jets.

\section{Conclusions}\label{sec:conclusions}

In high-energy heavy-ion collisions multiple hard processes take place in each collision event. We find that in central Pb-Pb collisions at the LHC energies a hard background produced by jet overlap effects starts to be a significant part of the underlying event for the jet reconstruction procedure. We have conducted a model study with initial state from EPOS3 model, which generates a complete initial configuration including soft and hard partons and has been proven to describe the soft and the hard sector of heavy ion collisions at LHC energies. This initial state for a central Pb-Pb collision contains a number of jet seeds - hard partons - with $p_\perp > 10$~GeV per event. Each initial hard parton is evolved from its initial high virtuality scale down to a lower virtuality scale $Q_0$ with Monte Carlo implementation of DGLAP equations. These jets evolve independently but can nevertheless overlap in momentum space. Standard jet finding algorithms like the (anti)-$k_\perp$ approach cannot distinguish between partons from the principal jets and those from neighbouring jets (intruders).

We obtained two major results:\\
1) the smaller the jet cone opening angle R (in the reconstruction algorithm) the less is the probability that intruder jets contribute to the momentum of the reconstructed jet but at the same time the smaller R the larger is the fraction of partons from the modelled (generator level) jet which are not clustered into the corresponding reconstructed jet. The latter is a known effect which results in a typically smaller $p_\perp$ of the reconstructed jets as compared to the underlying modelled jets.
The sum of both contributions, the gain of $p_\perp$ due to intruder partons and the loss of $p_\perp$ due to partons from the modelled jets which are outside of the jet cone can be negative or positive and depends on the jet opening angle R. Studying this effect as a function of R we find that there is no ``sweet spot'' which minimises both effects in central Pb-Pb collisions.\\
The relative effect of the intruders on the reconstructed jet $p_\perp$ decreases with the jet $p_\perp$, however:\\
2) The partons from the intruder jets modify the jet shape for jets with a $p_\perp$ as large as 80~GeV. The reason is that the $p_\perp$ of the principal jets is concentrated at low values of $r$ whereas the partons from the intruder jets contribute mostly to large $r$ values.

The jet overlap effect strongly depends on centrality: for mid-central (45-55\%) centrality class it becomes negligible.

Experimental analyses include elaborate techniques to subtract the jet overlap effects along with other background contributions. However such techniques are not standardised, not always clearly described and are to some extent model dependent, e.g.~PYTHIA jets are used as reference medium unmodified jets. Therefore, at the theory side one has to either simulate solitary jets passing through the QGP medium (which by construction does not introduce the overlap effect) or simulate full AA event which includes both multiple (mini-)jet production and low-momentum hadrons and apply a background removal technique similar to experimental one (which removes the overlap effect).

Here we limited our study to jets which do not interact with the medium. The influence of medium interactions on the jet shape and on the reconstructed $p_\perp$ will be subject of a future investigation.

\subsection*{Acknowledgements} We acknowledge enlightening discussions with Barbara Trzeciak and Matthew Nguyen. The work is supported by Region Pays de la Loire(France) under contract no.~2015-08473 and by the project Centre of Advanced Applied Sciences with the number: CZ.02.1.01/0.0/0.0/16-019/0000778. Project Centre of Advanced Applied Sciences is co-financed by European Union. M.~R.~acknowledges support by Polish National Science Center grant 2015/19/B/ST/00937.

\bibliographystyle{utphys}
\bibliography{refs-jets}

\providecommand{\href}[2]{#2}\begingroup\raggedright\begin{thebibliography}{10}

\bibitem{Shuryak:1977ut}
E.~V. Shuryak {\em Sov. Phys. JETP} {\bfseries 47} (1978) 212--219.
[Zh. Eksp. Teor. Fiz.74,408(1978)].

\bibitem{Gyulassy:1990ye}
M.~Gyulassy and M.~Plumer
\href{http://dx.doi.org/10.1016/0370-2693(90)91409-5}{{\em Phys. Lett.}
  {\bfseries B243} (1990) 432--438}.

\bibitem{Casalderrey-Solana:2015vaa}
J.~Casalderrey-Solana, D.~C. Gulhan, J.~G. Milhano, D.~Pablos, and K.~Rajagopal
\href{http://dx.doi.org/10.1007/JHEP03(2016)053}{{\em JHEP} {\bfseries 03}
  (2016) 053}.

\bibitem{KunnawalkamElayavalli:2017hxo}
R.~Kunnawalkam~Elayavalli and K.~C. Zapp
\href{http://dx.doi.org/10.1007/JHEP07(2017)141}{{\em JHEP} {\bfseries 07}
  (2017) 141}.

\bibitem{Sirunyan:2018qec}
{\bfseries CMS} Collaboration, A.~M. Sirunyan {\em et~al.}
\href{http://dx.doi.org/10.1103/PhysRevLett.121.242301}{{\em Phys. Rev. Lett.}
  {\bfseries 121} no.~24, (2018) 242301}.

\bibitem{Aaboud:2019oac}
{\bfseries ATLAS} Collaboration, M.~Aaboud {\em et~al.}
\href{http://dx.doi.org/10.1103/PhysRevLett.123.042001}{{\em Phys. Rev. Lett.}
  {\bfseries 123} no.~4, (2019) 042001}.

\bibitem{Becattini:2014hla}
F.~Becattini, E.~Grossi, M.~Bleicher, J.~Steinheimer, and R.~Stock
\href{http://dx.doi.org/10.1103/PhysRevC.90.054907}{{\em Phys. Rev.} {\bfseries
  C90} no.~5, (2014) 054907}.

\bibitem{Sharma:2018jqf}
N.~Sharma, J.~Cleymans, B.~Hippolyte, and M.~Paradza
\href{http://dx.doi.org/10.1103/PhysRevC.99.044914}{{\em Phys. Rev.} {\bfseries
  C99} no.~4, (2019) 044914}.

\bibitem{Dokshitzer:1997in}
Y.~L. Dokshitzer, G.~D. Leder, S.~Moretti, and B.~R. Webber
\href{http://dx.doi.org/10.1088/1126-6708/1997/08/001}{{\em JHEP} {\bfseries
  08} (1997) 001}.

\bibitem{Wobisch:1998wt}
M.~Wobisch and T.~Wengler in {\em {Monte Carlo generators for HERA physics.
  Proceedings, Workshop, Hamburg, Germany, 1998-1999}}, pp.~270--279.
\newblock
\href{http://arxiv.org/abs/hep-ph/9907280}{{\ttfamily arXiv:hep-ph/9907280
  [hep-ph]}}.
\newblock

\bibitem{Cacciari:2008gp}
M.~Cacciari, G.~P. Salam, and G.~Soyez
\href{http://dx.doi.org/10.1088/1126-6708/2008/04/063}{{\em JHEP} {\bfseries
  04} (2008) 063}.

\bibitem{Cacciari:2011ma}
M.~Cacciari, G.~P. Salam, and G.~Soyez
\href{http://dx.doi.org/10.1140/epjc/s10052-012-1896-2}{{\em Eur. Phys. J.}
  {\bfseries C72} (2012) 1896}.

\bibitem{Khachatryan:2016jfl}
{\bfseries CMS} Collaboration, V.~Khachatryan {\em et~al.}
\href{http://dx.doi.org/10.1103/PhysRevC.96.015202}{{\em Phys. Rev.} {\bfseries
  C96} no.~1, (2017) 015202}.

\bibitem{Adam:2015ewa}
{\bfseries ALICE} Collaboration, J.~Adam {\em et~al.}
\href{http://dx.doi.org/10.1016/j.physletb.2015.04.039}{{\em Phys. Lett.}
  {\bfseries B746} (2015) 1--14}.

\bibitem{Acharya:2018uvf}
{\bfseries ALICE} Collaboration, S.~Acharya {\em et~al.}
\href{http://dx.doi.org/10.1007/JHEP10(2018)139}{{\em JHEP} {\bfseries 10}
  (2018) 139}.

\bibitem{Sirunyan:2018jqr}
{\bfseries CMS} Collaboration, A.~M. Sirunyan {\em et~al.}
\href{http://dx.doi.org/10.1007/JHEP05(2018)006}{{\em JHEP} {\bfseries 05}
  (2018) 006}.

\bibitem{Acharya:2019ssy}
{\bfseries ALICE} Collaboration, S.~Acharya {\em et~al.}
\href{http://dx.doi.org/10.1016/j.physletb.2019.07.020}{{\em Phys. Lett.}
  {\bfseries B796} (2019) 204--219}.

\bibitem{Werner:2013tya}
K.~Werner, B.~Guiot, I.~Karpenko, and T.~Pierog
\href{http://dx.doi.org/10.1103/PhysRevC.89.064903}{{\em Phys. Rev.} {\bfseries
  C89} no.~6, (2014) 064903}.

\bibitem{Drescher:2000ha}
H.~J. Drescher, M.~Hladik, S.~Ostapchenko, T.~Pierog, and K.~Werner
\href{http://dx.doi.org/10.1016/S0370-1573(00)00122-8}{{\em Phys. Rept.}
  {\bfseries 350} (2001) 93--289}.

\bibitem{Werner:2007bf}
K.~Werner
\href{http://dx.doi.org/10.1103/PhysRevLett.98.152301}{{\em Phys. Rev. Lett.}
  {\bfseries 98} (2007) 152301}.

\bibitem{Werner:2010aa}
K.~Werner, I.~Karpenko, T.~Pierog, M.~Bleicher, and K.~Mikhailov
\href{http://dx.doi.org/10.1103/PhysRevC.82.044904}{{\em Phys. Rev.} {\bfseries
  C82} (2010) 044904}.

\bibitem{Rohrmoser:2018fkf}
M.~Rohrmoser, P.-B. Gossiaux, T.~Gousset, and J.~Aichelin
\href{http://dx.doi.org/10.5506/APhysPolB.49.1325}{{\em Acta Phys. Polon.}
  {\bfseries B49} (2018) 1325}.

\bibitem{Acharya:2017goa}
{\bfseries ALICE} Collaboration, S.~Acharya {\em et~al.}
\href{http://dx.doi.org/10.1016/j.physletb.2017.11.044}{{\em Phys. Lett.}
  {\bfseries B776} (2018) 249--264}.

\bibitem{Aad:2019igg}
{\bfseries ATLAS} Collaboration, G.~Aad {\em et~al.}
\href{http://dx.doi.org/10.1103/PhysRevC.100.064901}{{\em Phys. Rev.}
  {\bfseries C100} no.~6, (2019) 064901}.

\bibitem{Tachibana:2017syd}
Y.~Tachibana, N.-B. Chang, and G.-Y. Qin
\href{http://dx.doi.org/10.1103/PhysRevC.95.044909}{{\em Phys. Rev.} {\bfseries
  C95} no.~4, (2017) 044909}.

\bibitem{Putschke:2019yrg}
J.~H. Putschke {\em et~al.}
{\em arXiv} (2019) 1903.07706.

\bibitem{He:2018xjv}
Y.~He, S.~Cao, W.~Chen, T.~Luo, L.-G. Pang, and X.-N. Wang
\href{http://dx.doi.org/10.1103/PhysRevC.99.054911}{{\em Phys. Rev.} {\bfseries
  C99} no.~5, (2019) 054911}.

\end{thebibliography}\endgroup

\end{document}